%% file: main.tex
\documentclass[conference]{IEEEtran}

\usepackage{graphicx}
\usepackage{soul}
\usepackage{xcolor}
\usepackage[inkscapelatex=false]{svg}
\usepackage{listings}
\usepackage{comment}
\usepackage{float}
\usepackage[hidelinks]{hyperref}
\usepackage{tabularx}
\usepackage{booktabs}
\usepackage{multirow} 
\usepackage{graphicx} 
\usepackage{adjustbox} 
\usepackage{colortbl} \usepackage[table,xcdraw]{xcolor}
\usepackage[T1]{fontenc}
\usepackage[noadjust]{cite}
\usepackage[ruled, vlined, longend]{algorithm2e}
\usepackage{algpseudocode}

\SetCommentSty{mycommfont}

\newcommand{\PreserveBackslash}[1]{\let\temp=\\#1\let\\=\temp}
\newcolumntype{C}[1]{>{\PreserveBackslash\centering}p{#1}}
\newcolumntype{R}[1]{>{\PreserveBackslash\raggedleft}p{#1}}
\newcolumntype{L}[1]{>{\PreserveBackslash\raggedright}p{#1}}
\newcolumntype{P}[1]{>{\centering\arraybackslash}p{#1}}

\usepackage{caption}
\makeatletter
\newcommand\notsotiny{\@setfontsize\notsotiny\@viipt\@viipt}
\makeatother

\newcommand{\DejaVu}{Déjà Vu}
\newcommand{\dejavu}{déjà vu}

\definecolor{codegreen}{rgb}{0,0.6,0}
\definecolor{codegray}{rgb}{0.5,0.5,0.5}
\definecolor{codepurple}{rgb}{0.58,0,0.82}
\definecolor{backcolour}{rgb}{0.95,0.95,0.92}

\sethlcolor{yellow}

\begin{document}

\title{
\DejaVu~Packing: Optimizing FPGA Logic Clustering Runtime via Pattern Memoization
\vspace{-0.3cm}
}

\author{
\IEEEauthorblockN{
    Milo Liebster\textsuperscript{\textdagger}, 
    Amin Mohaghegh\textsuperscript{*}, 
    Andrew Boutros\textsuperscript{\textdagger}}
\IEEEauthorblockA{
\textsuperscript{\textdagger}Department of Electrical and Computer Engineering, University of Waterloo \\
\textsuperscript{*}QuickLogic Corporation\\
Email: \{pliebste, andrew.boutros\}@uwaterloo.ca, amohaghegh@quicklogic.com}
\vspace{-0.9cm}
}

\maketitle

\begin{abstract}
Implementing a digital circuit on a field-programmable gate array (FPGA) fabric requires clustering technology-mapped netlist primitives into coarser-granularity blocks that can be directly mapped to the physical resources available on the FPGA fabric.
As the internal architecture of FPGA logic blocks (LBs) has grown in complexity, with sophisticated logic elements (LEs) and highly irregular local interconnect, this packing problem has become significantly more challenging. 
To ensure the feasibility of intracluster routing, the computer-aided design (CAD) tools must solve a costly multi-source multi-sink routing problem for each candidate cluster.
In this paper, we first show that such packing legality checks consume a significant portion of the CAD flow runtime for LB architectures with complex LEs and local routing structures resembling modern commercial FPGAs.
We demonstrate that the packing stage constitutes 58\% and 94\% of the entire Versatile Place and Route (VPR) flow runtime on average when mapping a wide variety of benchmarks to the AMD 7-series-like and Altera Stratix 10--like VTR architecture captures, respectively.
By analyzing the packing algorithm behavior, we observe that a significant fraction of the attempted packed clusters are repetitions of a much smaller number of packing patterns, and therefore many of the packing legality checks are redundant and could be skipped. 
To this end, we introduce our \DejaVu~packing approach, which leverages a novel packing signature tree data structure that enables efficient identification of recurring packing patterns and memoization of their legality check outcomes.
Our approach speeds up the packing runtime by up to 13.4$\times$ and 29.3$\times$, with an average of 3.7$\times$ and 6.9$\times$, across the evaluated benchmarks on the 7-series and Stratix 10 architecture captures.
These packing runtime gains result in a significant 1.6$\times$ and 5.3$\times$ average reduction in end-to-end VPR runtime, while maintaining quality of results.
\end{abstract}

\IEEEpeerreviewmaketitle

\fontsize{10pt}{10.696pt}\selectfont

\input{01-introduction}
\input{02-background}
\input{03-characterization}
\input{04-pst}
\input{05-dejavu-packing}
\input{06-results}
\input{07-conclusion}

\section*{Acknowledgment}
This work was funded by the Natural Sciences and Engineering Research Council of Canada (NSERC) and the University of Waterloo President's Research Award. The authors would like to thank Alexandre Singer, Amirhossein Poolad, and Vaughn Betz for their insightful feedback during integrating this work into the VTR master branch.

\bibliographystyle{IEEEtran}
\bibliography{references}

\end{document}

%% file: 01-introduction.tex
\section{Introduction}
\label{sec:intro}

Field-programmable gate array (FPGA) development relies on a multi-stage computer-aided design (CAD) flow that maps descriptions of digital circuits onto the configurable fabric.
Synthesis and technology mapping first translate the design logic into a netlist of \emph{primitives} available on the FPGA, such as lookup tables (LUTs), flip-flops (FFs), adders, multipliers, and memory elements.
Then, an explicit packing stage can be used to organize these primitives, based on their types, into coarser-granularity \emph{clusters} that can be mapped to the \emph{physical blocks} in the target FPGA architecture, such as logic blocks (LBs), digital signal processing blocks, and block RAMs.
Poor packing decisions at this step directly impact downstream placement and routing quality, resulting in increased wire congestion, longer critical paths, and reduced overall circuit performance. 
Alternatively, some contemporary CAD flows first perform an initial placement of netlist primitives to guide subsequent packing decisions or directly place netlist primitives at specific cluster locations and iteratively refine any invalid or sub-optimal cluster mappings~\cite{li2019paradigm}.

Among the various block types in modern FPGAs, LBs are the most challenging to cluster due to their architectural complexity and heterogeneity.
They contain tens of LUTs, FFs, and hardened adders~\cite{murray2020optimizing}, all interconnected through a complex and highly irregular local routing network.
Unlike simpler block types with tighter constraints, modern LB architectures offer numerous possible configurations for clustering primitives, but not all clusters are physically realizable (i.e., \emph{legal}).
The CAD flow must ensure that each candidate cluster has a valid intracluster routing solution.
This involves costly legality checks that require solving multi-source, multi-sink routing problems within the constrained local interconnect graph.
Since many candidate configurations may be attempted per cluster (in an explicit packing stage or during flat placement legalization), these legality checks are performed thousands or even millions of times, consuming substantial runtime.

By analyzing the runtime of the widely used open-source Versatile Place and Route (VPR) \cite{elgammal2025vtr9} tool, we demonstrate that cluster legality checks during the packing stage constitute the majority of the flow runtime for architectures that resemble modern commercial FPGAs with complex LBs.
However, our analysis also reveals that the runtime spent checking cluster legality is dominated by redundant checks, assessing the legality of recurring cluster patterns that have been previously encountered and evaluated during packing.
Based on this observation, we present a novel \emph{packing signature tree} (PST) data structure that enables efficient memoization and comparison of cluster packing patterns by assigning them unique representations encoded as paths through a tree, which we refer to as \emph{packing signatures}.
Then, we integrate our PST into VPR's packing algorithm to characterize a wide variety of circuits from the Verilog-to-Routing (VTR), Koios~\cite{arora2023koios}, Titan23~\cite{murray2013titan}, and Titanium25~\cite{elgammal2025vtr9} benchmark suites.
Our experiments target both the AMD 7-series and Altera Stratix 10 VTR architecture captures.
However, we modify the VTR Stratix 10 capture to more accurately model the complex LB internals of the commercial Stratix 10 device.

We find that, on average, 62\% of the clusters subjected to legality checks are repeats of clusters that have been previously evaluated for legality. 
We show that by memoizing the legality check outcomes of these clusters within our PST, enabling redundant checks to be skipped, the VPR packing runtime can be reduced by up to 29.3$\times$.
This approach speeds up the entire VPR flow by an average of 1.6$\times$ and 5.3$\times$ across 68 benchmarks on the 7-series and Stratix 10 architectures, respectively.
Our proposed memoization strategy does not involve any heuristics;
therefore, it results in identical packing solutions, and thus quality of results (i.e., routed wirelength and critical path delay), to those produced by the unmodified flow.
In summary, our contributions include:
\begin{itemize}
     \item Demonstrating that the majority of clusters produced by greedy seed-based packing recur from a much smaller set of distinct packing patterns.
     \item Introducing a novel packing signature tree data structure to memoize and enable efficient recognition of recurrent packing patterns and their legality check outcomes.
     \item Developing a Stratix 10--like VTR architecture file that faithfully captures the complexity of the commercial Stratix 10 LB architecture.
     \item Reducing VPR's packing and end-to-end runtime by 5.1$\times$ and 2.9$\times$ on average across a wide variety of benchmarks and two architecture captures of modern commercial devices with complex LBs.
\end{itemize}

Our packing memoization technique is integrated into VTR and can be configured using the command-line argument \mbox{\texttt{---memoize\_cluster\_packings \{on|off\}}}.

%% file: 02-background.tex
\section{Background \& Related Work}
\label{sec:background}

\subsection{FPGA Packing}

Most of the prior work on FPGA packing algorithms focuses on seed-based approaches in which the packer iteratively selects a \emph{seed} primitive to form a new cluster and then greedily adds related primitives until the cluster is full or no more legal additions can be made.
One of the earliest FPGA seed-based packers is VPack~\cite{betz1997cluster}, which aims to maximize cluster utilization through simple heuristics. 
The unpacked primitive with the most used pins is selected as a cluster seed, and subsequent primitives are greedily added to the cluster based on their connectivity to already-packed elements.
\mbox{T-VPack}~\cite{marquardt1999using} extends VPack by making it timing-driven.
It successively clusters primitives along the circuit's critical path to minimize the number of external connections between them and re-evaluates primitive criticality after each decision.

Other seed-based packers explore alternative optimization objectives and clustering strategies. 
For example, RPack~\cite{bozorgzadeh2001rpack} uses a packing cost function that aims to improve routability, while iRAC~\cite{singh2002efficient} improves routing congestion by trying to match the Rent exponent of the clustered design to that of the underlying FPGA architecture.
HD-Pack~\cite{chen2007improving} performs a rapid global placement to determine approximate primitive locations and uses this information to better inform packing decisions. 
MO-Pack~\cite{rajavel2011mo} integrates ideas from multiple prior works to co-optimize energy, critical path delay, and channel width concurrently using a many-objective packing algorithm. 
Depopulation-based packers, such as Un/DoPack~\cite{tom2006dopack} and T-NDPack~\cite{liu2009t}, improve routing congestion by deliberately producing partially filled clusters. 
This technique spreads out the circuit across a larger area of the fabric and therefore reduces the peak utilization of routing channels.

Besides seed-based approaches, several prior works explore partitioning-based packing algorithms that split the netlist into an initial set of clusters using a k-way partitioner, such as hMetis~\cite{karypis1999multilevel}, and then perform a legalization step to fix these clusters. 
PPack~\cite{feng2012k, feng2014rent} uses recursive bipartitioning to optimize the Rent characteristic of the cluster-level netlist instead of simply minimizing the number of external signals. 
More recent partitioning approaches such as PartSA~\cite{vercruyce2016runtime} and MultiPart~\cite{vercruyce2017preserving} have incorporated multithreading to improve packing runtime. 
Singhal et al.~\cite{singhal2017lsc} present a consensus-based packing technique in which each primitive independently creates candidate clusters and iteratively reaches consensus with other members of the cluster.
This approach mitigates the greedy behavior of seed-based packers while enabling multithreaded clustering since many primitives can concurrently form their candidate clusters and evaluate their legality. 
Since accurate and fast estimates of timing, wirelength, and congestion of the placed netlist are not available during the packing stage, some prior work~\cite{chen2004simultaneous,elgammal2023breaking} allow packed primitives to retroactively be moved and swapped between clusters during placement to further improve the final implementation quality of results. 

These partitioning-based and hybrid methods have not achieved widespread adoption in academic tools, largely due to their reduced adaptability across different FPGA architectures and heterogeneous blocks compared to seed-based approaches.
The current academic state-of-the-art for seed-based packing is Architecture-Aware Packing or AAPack~\cite{Luu2011AAPack, luu2014towards} developed for the VTR~\cite{elgammal2025vtr9} open-source FPGA toolflow.
VTR takes as an input an XML-based file that describes the details of an FPGA architecture, such as its block types, organization, timing/area models, and routing resources between and within blocks.
The flow consists of multiple tools that synthesize benchmark circuits, map them to the specified FPGA architecture, and report implementation results such as resource utilization, critical path delay, and routing wirelength. 

To enable flexible architecture exploration in VTR, AAPack can target any arbitrary logic block architecture defined using VTR's architecture description format with features such as fracturable LUTs, hardened arithmetic, and depopulated crossbars~\cite{lemieux2001using}. 
Supporting complex and arbitrary architectures has significantly increased the computational cost of cluster legality checking, as verifying routing feasibility can require solving numerous multi-source, multi-sink intracluster routing problems to form even a single cluster.
AAPack introduced speculative packing to mitigate this overhead by optimistically skipping all intermediate legality checks in its first attempt to pack a cluster; however, as LB complexity increases, the savings achieved by this approach are reduced since the probability of packing a legal cluster without intermediate legality checks becomes significantly lower.
To further address packing runtime, RSVPack pre-computes routing feasibility tables representing all legal cluster routing configurations of a Virtex~6 LB~\cite{haroldsen2017academic}.
The table is efficiently iterated through to assess cluster legality, achieving an average 25$\times$ speedup compared to AAPack. 
However, it is not described whether RSVPack can evaluate a solution space involving different permutations of logically equivalent pins (i.e. interchangable cluster and LUT inputs connected by a crossbar), thus it is unclear if this method could be generalized for arbitrary architectures.
To provide a general method for reducing packing runtime, our work introduces a lazy method for memoizing the results of the legality checks of previously encountered packing patterns. 
This eliminates redundant legality checks resulting in significant runtime gains, especially for architectures with complex LB structures.
Although we showcase our proposed technique for the seed-based packing algorithm in VPR, we believe it is also applicable to other packing algorithms and CAD flows in which intracluster routing (or other costly checks) must be performed to guarantee legality.

\subsection{Architecture Aware Packing Algorithm in VPR}
\label{sec:vpr_packing}

Algorithm~\ref{alg:vpr-packer} summarizes the packing algorithm currently used in VPR, which was originally introduced in~\cite{Luu2011AAPack} and then refined over several VTR releases~\cite{Luu2014vtr7,murray2020vtr8,elgammal2025vtr9} to improve its runtime and quality of results (QoR).
The input to the packing stage is a technology-mapped netlist of interconnected primitives, such as lookup tables (LUTs), flip-flops (FFs), memories, adders, and multipliers.
First, a pre-packing step is performed which groups adjacent netlist primitives (also called \emph{atoms}) that have highly-constrained placement relationships relative to each other into \emph{molecules} whose atoms must be placed together in fixed arrangements during packing.
Examples of such molecules are LUTs directly driving FFs, or adders connected via dedicated carry chains~\cite{Luu2014vtr7}.
Primitives that do not have such constraints are pre-packed into single-atom molecules.

\setlength{\textfloatsep}{5pt}
\begin{algorithm}[t!]
    \caption{Overview of the VPR packing algorithm.}
    \label{alg:vpr-packer}
    \SetAlgoVlined
    
    \SetKwProg{Fn}{Function}{ is}{end}
    \SetKw{Continue}{continue}
    \Fn{Pack(netlist)}{
        \While{netlist.has\_unpacked\_molecules()}{
            \textit{seed\_molecule} $\gets$ \textit{netlist.get\_next\_seed()}\;
            \textit{cluster} $\gets$ \textit{start\_cluster(seed\_molecule)}\;

            \tcc{Speculative Packing}
            \While{cluster.capacity() < max\_util}{
                \textit{molecule} $\gets$ \textit{netlist.get\_next\_mol(cluster)}\;
                \If{cluster.check\_pin\_count(molecule)}{
                    \textit{cluster.insert(molecule)}\;
                }
            }
            \If{cluster.check\_routability()}{
                \textit{cluster.finalize()}\;
                \Continue\;
            }

            \tcc{Detailed Packing}
            \textit{cluster.reset()}\;
            
            \While{cluster.capacity() < max\_util}{
                \textit{molecule} $\gets$ \textit{netlist.get\_next\_mol(cluster)}\;
                \If{cluster.check\_pin\_count(molecule)}{
                    \textit{cluster.insert(molecule)}\;
                    \If{!cluster.check\_routability()}{
                        \textit{cluster.remove(molecule)}\;
                        \Continue;
                    }
                }
            }
            \textit{cluster.finalize()}\;
        }
    }
\end{algorithm}

The packing algorithm begins by picking a seed molecule according to a gain function parameterized by pin utilization and timing criticality and placing it into a new cluster of the appropriate type (e.g., LB, BRAM, DSP) to accommodate it.
Following seed placement, the algorithm continues to greedily select the most-related unpacked molecules that can be inserted into the open cluster, based on an empirically derived \emph{attraction function}, until the user-recommended pin utilization is achieved or all candidate molecules have been exhausted.
The default VPR packer settings set the target input pin utilization for LBs to 80\% and disable packing of unrelated molecules, which was shown to achieve better QoR~\cite{murray2020vtr8}.

Since VTR allows the description of arbitrary block primitives and local interconnect structures, it is necessary to ensure that the intracluster routing among the molecules in a packed cluster and its external pins is feasible.
The VPR packer first performs \emph{speculative packing} in which only a simple pin counting legality check is performed each time a molecule is added to a cluster to rapidly discard packing solutions that exceed the number of available cluster pins.
Then, only when a cluster is complete, the intracluster router is invoked to check its routing feasibility.
If the speculative packing attempt fails the routing legality check, the cluster is reset and a \emph{detailed packing} is performed where the intracluster router is invoked after the insertion of each molecule.
This continues until the cluster is filled to the desired packing density or until the cluster becomes unroutable if any of the remaining unpacked molecules is inserted.

\begin{figure}[t!]
    \centering{\includegraphics[width=1\columnwidth]{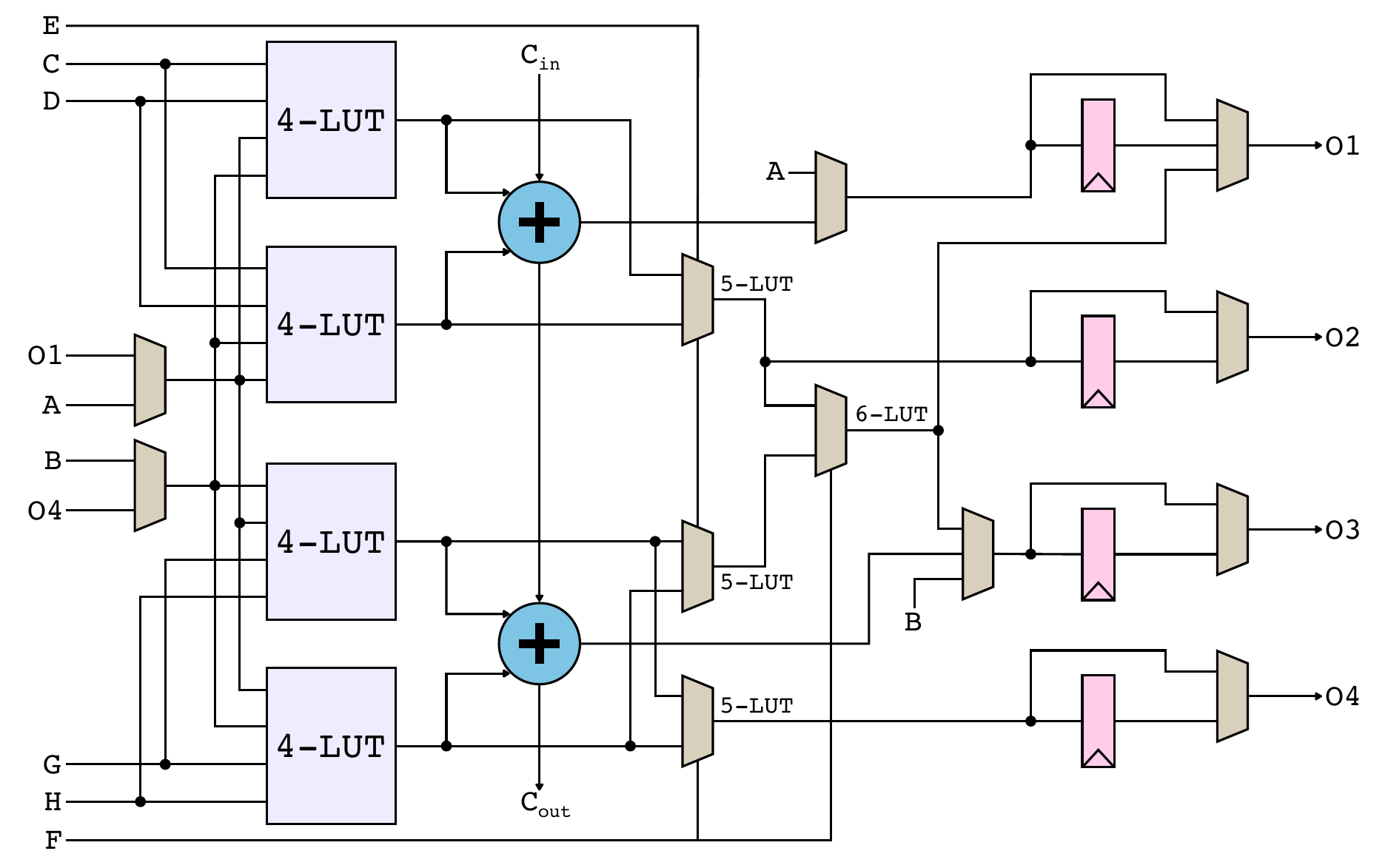}}
    \caption{Complex commercial Stratix 10 LE with 8 distinct inputs (A-H), one fracturable 6-LUT, two bits of arithmetic, and four bypassable FFs. Finding a packing solution with feasible routing can be computationally expensive for this LB organization since not all inputs can feed both 5-LUTs in fractured mode, only 2 registers (O1, O3) can be directly reached from 2 specific inputs (A, B), and adder outputs can only feed specific registers (O1, O3).}
    \label{fig:s10_alm}
\end{figure}

\begin{figure*}
    \centering{\includegraphics[width=2\columnwidth]{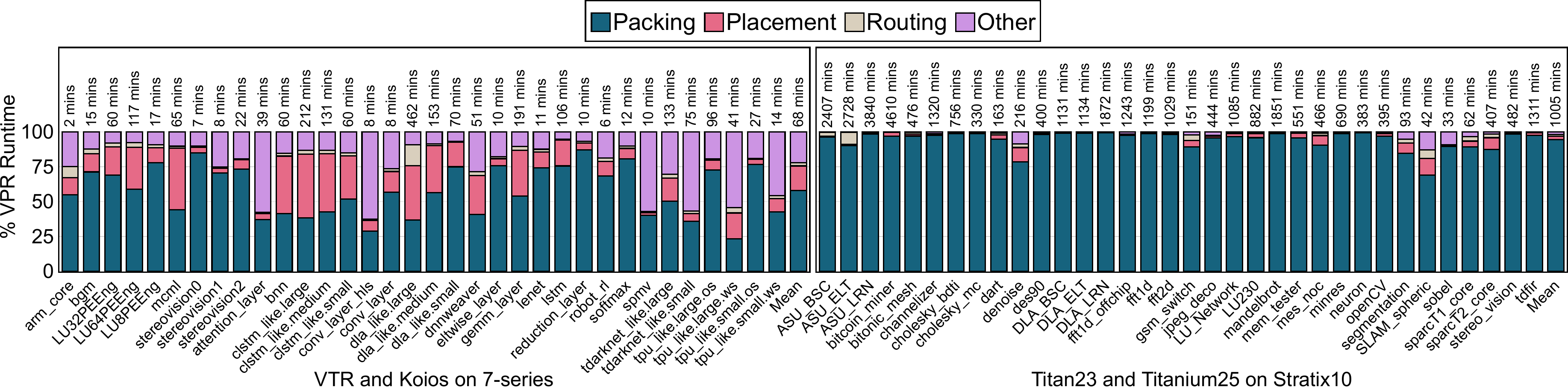}}
    \caption{VPR runtime breakdown for the VTR and Koios benchmarks on the 7-series architecture (W=300) (left), as well as the Titan23 and Titanium25 benchmarks on Stratix 10 architecture (W=400) (right). On average, the packing runtime is 58\% (39 mins) and 94\% (945 mins) of the overall VPR runtime (68 and 1005 mins).}
    \label{fig:baseline-runtimes}
    \vspace{-0.5cm}
\end{figure*}

\subsection{7-series \& Stratix 10 VTR Architecture Captures}
\label{sec:vtr_arch}

The latest VTR release~\cite{elgammal2025vtr9} includes approximate models of the AMD 7-series and the Altera Stratix 10 commercial architectures.
The LB in the 7-series architecture has 56 inputs and 24 outputs split between two slices.
A slice contains four logic elements (LEs), each of which has a six-input fracturable LUT (6-FLUT), two bypassable FFs, and one bit of arithmetic.
The local interconnect drives each slice input from one LB input or one of three slice outputs (i.e., feedback connections), one or two of which are produced by the same slice.
The two FFs in an LE can be driven by the 6-FLUT or directly reached through a specific LE input, skipping the LUTs.
Additionally, one of the two FFs can be driven by the output of the hard arithmetic circuitry.

Talaei et al.~\cite{khoozani2023titan} introduced the VTR Stratix 10 architecture capture along with an updated Titan flow~\cite{murray2013titan} to support newer Altera FPGA families. 
The Titan flow enables synthesizing benchmarks using the Altera Quartus CAD tool to produce circuit netlists in \texttt{blif} format, which can then be packed, placed, and routed using VPR.
In this architecture capture, 10 LEs are grouped into an LB with a total of 60 inputs and a full local crossbar. 
Each LE has eight distinct inputs that can drive \emph{any} of the six inputs of an FLUT-6, two bypassable FFs that can optionally be directly driven by \emph{any} LE input (skipping the LUTs), and two bits of arithmetic.
This is an overly simplified model compared to the commercial Stratix 10 LE illustrated in Fig.~\ref{fig:s10_alm}.
A faithful model of the Stratix 10 architecture would instead have a 50\% depopulated crossbar, a more restrictive connectivity pattern between the eight LE inputs and six FLUT inputs, and four FFs per LE (two of which can be  directly reached each by a single LE input, skipping the LUTs).
Based on communication with the authors of~\cite{khoozani2023titan}, these complex details of the Stratix 10 LB architecture were omitted as they resulted in a significant increase in the VPR packing runtime, prohibiting effective architecture and CAD algorithm evaluations.
Our work solves this issue and thus enables exploration of complex LB architectures similar to those in modern commercial FPGAs.

%% file: 03-characterization.tex
\section{Baseline Packing Results}
\label{sec:baseline_packing}

\subsection{Faithful Stratix 10 LB Architecture Model}
\label{sec:s10_modified}

To characterize the VPR packing algorithm on realistic LB architectures representative of commercial devices, we first modify the original VTR Stratix 10 architecture capture~\cite{khoozani2023titan} to incorporate the omitted details mentioned in Section~\ref{sec:vtr_arch}.
We incorporate a 50\% depopulated local crossbar built out of smaller full crossbars.
The LB inputs are divided into four groups of 15 logically equivalent pins where each of the eight inputs to an LE can be driven by any signal from two of the four groups.
We also increase the number of FFs from two to four, and capture the details of the complex connectivity between LE primitives as illustrated in Fig.~\ref{fig:s10_alm}.
Our modified architecture file has been contributed to the open-source VTR repository and will be referred to as the Stratix 10 architecture throughout the experiments described in the rest of this paper.

\subsection{VPR Packing Characterization}
\label{sec:vpr_char}

In this subsection, we characterize the behavior and runtime of the VPR packer for circuits with more than $10,000$ netlist primitives from the VTR and Koios~\cite{arora2023koios} benchmark suites mapped to the 7-series architecture as well as the Titan23~\cite{murray2013titan} and Titanium25~\cite{elgammal2025vtr9} benchmark suites mapped to the Stratix 10 architecture\footnote{BLIFs of the Titan23 and Titanium25 benchmarks are sourced from the Titan flow which uses Quartus as its synthesis frontend, and therefore are only compatible with the VTR Stratix IV and Stratix 10 architecture captures. Conversely, evaluating the VTR and Koios benchmarks on the VTR Stratix 10 capture would require migrating and synthesizing them for Stratix 10 in Quartus to obtain their VTR-compatible BLIFs, which we believe is out of the scope of this work.}.
We add telemetry code to analyze the main contributors to packing runtime in the VTR master branch code at the time we started this work (commit ID: \texttt{7c0ee8a6}).
All the reported results are averages of running VTR with 3 seeds on an Intel Xeon w9-3495 CPU with 1 TB of RAM.

Fig.~\ref{fig:baseline-runtimes} presents a breakdown of the end-to-end VPR runtime for different benchmarks and architectures.
The 7-series results show that the packing stage constitutes 58\% (about 39 mins) of the entire flow runtime on average.
The packing runtime increases to 94\% (945 mins) of the VPR runtime when mapping the Titan23 and Titanium25 benchmarks on the more complex Stratix 10 architecture. 
To put these results in perspective, when mapping the VTR and Koios benchmarks to the VTR flagship architecture with a full local LB crossbar, the packing stage constitutes only 29\% (3 mins) of the end-to-end VPR runtime (9 mins) on average.
This clearly highlights the significant impact of modeling complex LB local interconnect and LE structures on packing runtime, which hinders further exploration of such architectures that resemble modern commercial devices in VPR.

\begin{figure}[t]
    \centering{\includegraphics[width=1\columnwidth]{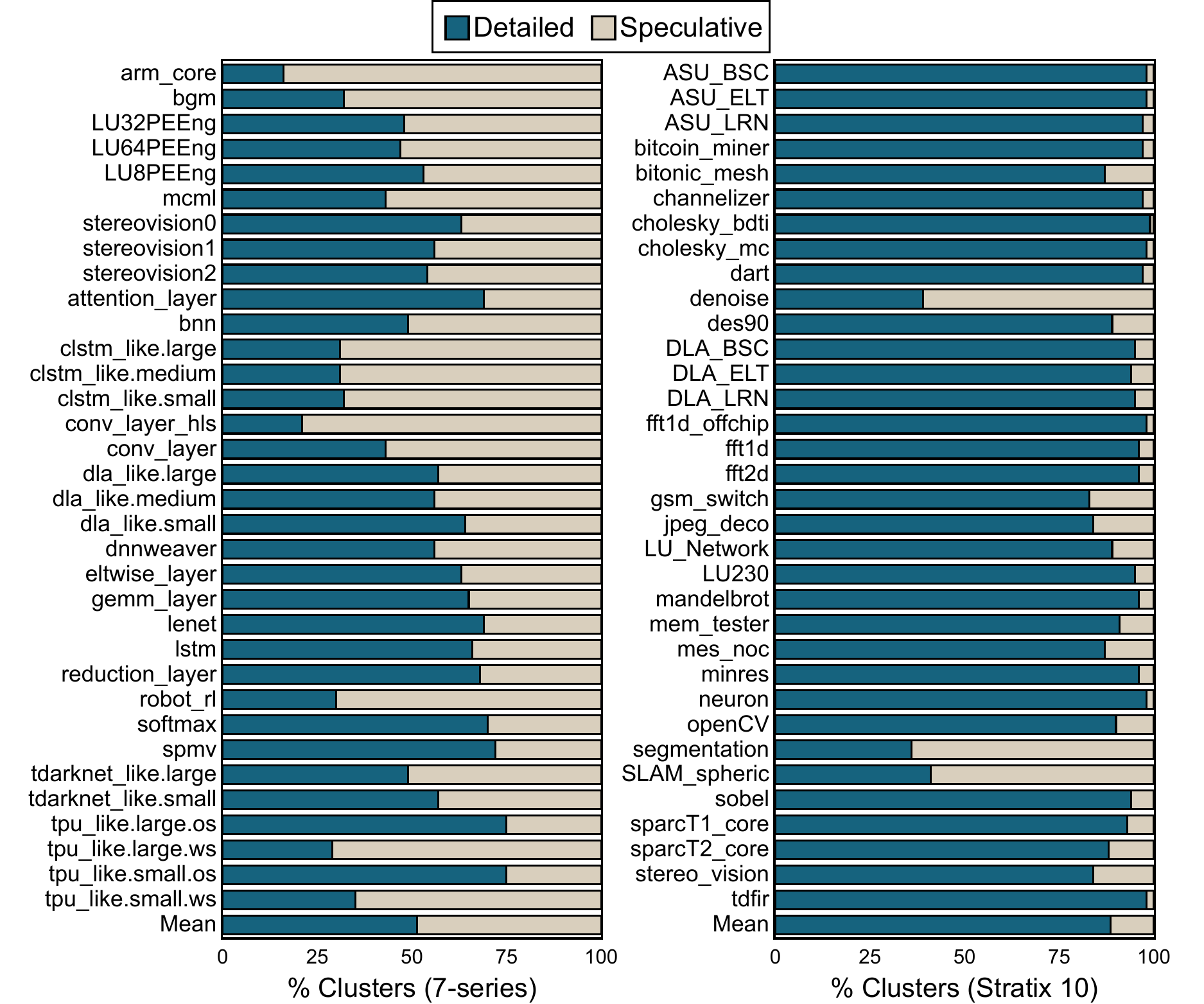}}
    \vspace{-0.6cm}
    \caption{Breakdown of the percentage of clusters that succeeded with speculative packing vs. clusters that required detailed packing. On average, 51\% and 89\% of the clusters required detailed packing in the case of VTR and Koios benchmarks on 7-series (left) and Titan32 and Titanium25 benchmarks on Stratix 10 (right), respectively.}
    \label{fig:detailed-packing}
\end{figure}

\begin{figure}[t]
    \centering{\includegraphics[width=1\columnwidth]{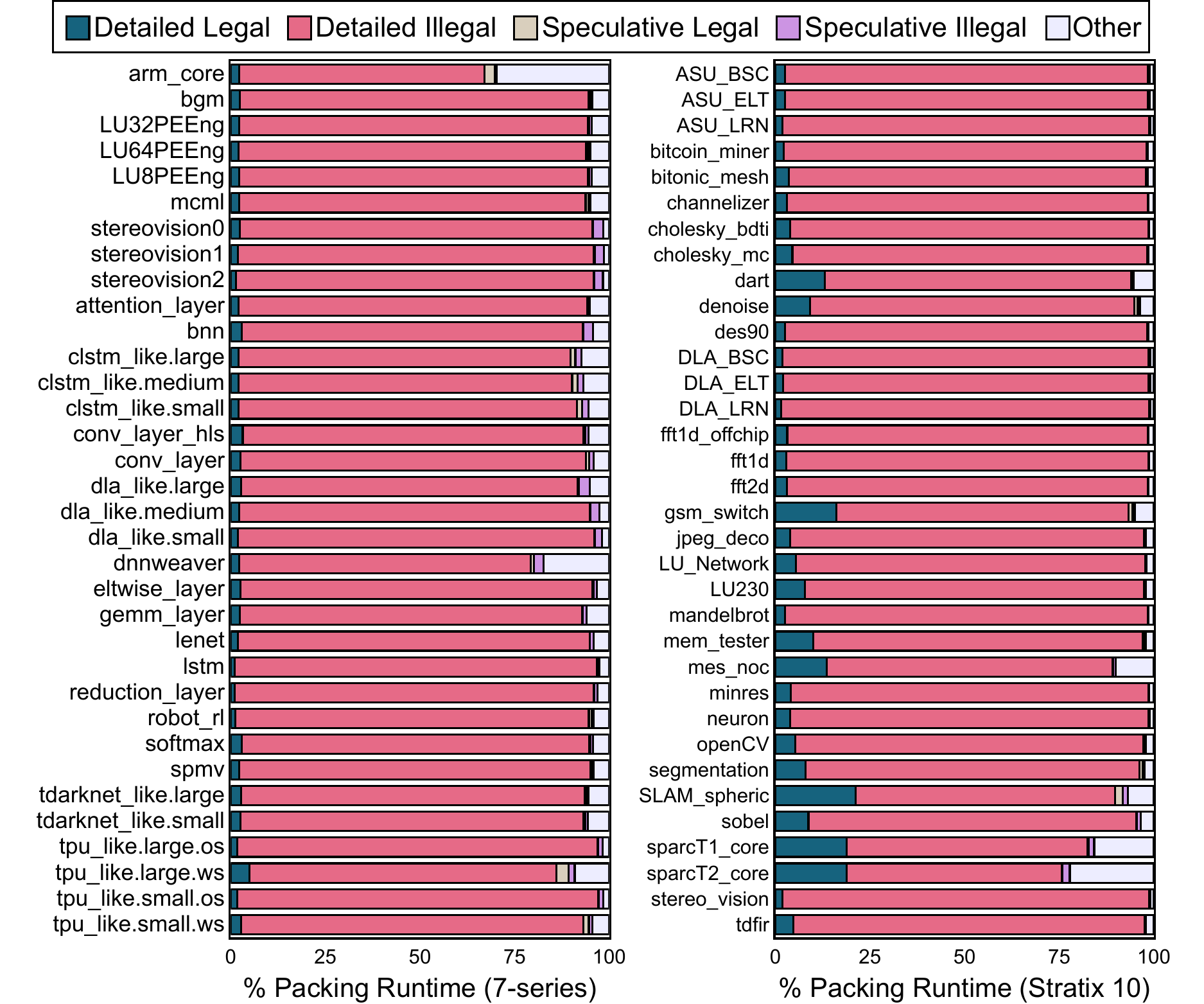}}
    \caption{Breakdown of the the packing runtime for the VTR and Koios benchmarks on the 7-series architecture (left) and Titan23 benchmarks on the Stratix 10 architecture (right). On average, 91\% and 90\% of the packing runtime is spent on intracluster routing of failed detailed packing attempts.}
    \label{fig:packing_breakdown}
\end{figure}

To understand the reason for this significant runtime increase, we analyze the packing behavior of all the circuits we experiment with to determine the percentage of logic clusters for which speculative packing (see Section~\ref{sec:vpr_packing}) failed to find a legal solution. Fig.~\ref{fig:detailed-packing} shows that 51\% and 89\% of all clusters failed the speculative packing step and resorted to detailed packing to find a legal solution for the 7-series and Stratix 10 architectures, respectively.
The Stratix 10 architecture generally shows a smaller percentage of speculative packing success due to its more complicated LB compared to the 7-series architecture, emphasizing that speculative packing becomes less effective as LB architecture increases in complexity.

Fig.~\ref{fig:packing_breakdown} presents the breakdown of packing runtimes into the time spent on failed and successful attempts of speculative and detailed packing.
It shows that, on average, 90\% of the packing runtime is attributed to the detailed packing step failing to route clusters, and partial clusters, that have no feasible routing solution. 
Note that even for benchmarks where clusters that fail speculative packing make up the minority (such as \texttt{arm\_core} in Fig.~\ref{fig:detailed-packing}), detailed packing still overshadows speculative packing in terms of total runtime (as seen in Fig.~\ref{fig:packing_breakdown}). 
The reason for this is that while successful speculative packing must only route a cluster once, clusters that fall back to detailed packing will commonly require tens or hundreds of routing calls to reach a solution on the 7-series architecture; 
this estimate increases by one to two orders of magnitude for Stratix 10. 
In the absolute worst case, individual clusters in our experiments have seen intracluster router invocation counts up to $2,254$ and $12,133$ on 7-series and Stratix 10 architectures, respectively.
These results indicate that significant runtime gains could be realized if the number of intracluster routing attempts made for these unroutable cluster candidates is reduced.
In Section~\ref{sec:pst}, we demonstrate that the majority of these problematic cluster patterns are repeats of previously evaluated packing signatures.

%% file: 04-pst.tex
\section{Packing Signature Tree}
\label{sec:pst}

\subsection{Data Structure Overview}
\label{sec:pst_ds}

To memoize LB packing patterns, we require an approach for comparing graphs of partially or completely packed clusters for equivalency. 
In its general form, this is a computationally complex graph isomorphism problem;
however, our PST data structure leverages properties of greedy packing algorithms to create tree-based packing signature encodings that can be compared in roughly constant time. 

The deterministic greedy behavior of common packing algorithms means that repetitions of identical substructures within a technology-mapped netlist (which we postulate are abundant in hierarchical digital circuits) will have equivalent affinity metrics computed for their constituent molecules. 
Consequently, we expect that such identical substructures will receive identical treatment during packing, resulting in equivalent packing patterns. 
We also assume that a consistent ordering is imposed on the insertion of molecules into equivalent clusters.
Under these conditions, which we verify empirically in the next subsection, detecting packing pattern equivalency can be implemented efficiently as traversal over a tree data structure where each node describes a packed primitive and its connectivity. 
We define each path from the tree root to a leaf node to represent a unique packing signature.

\begin{figure}
    \centering{\includegraphics[width=0.95\columnwidth]{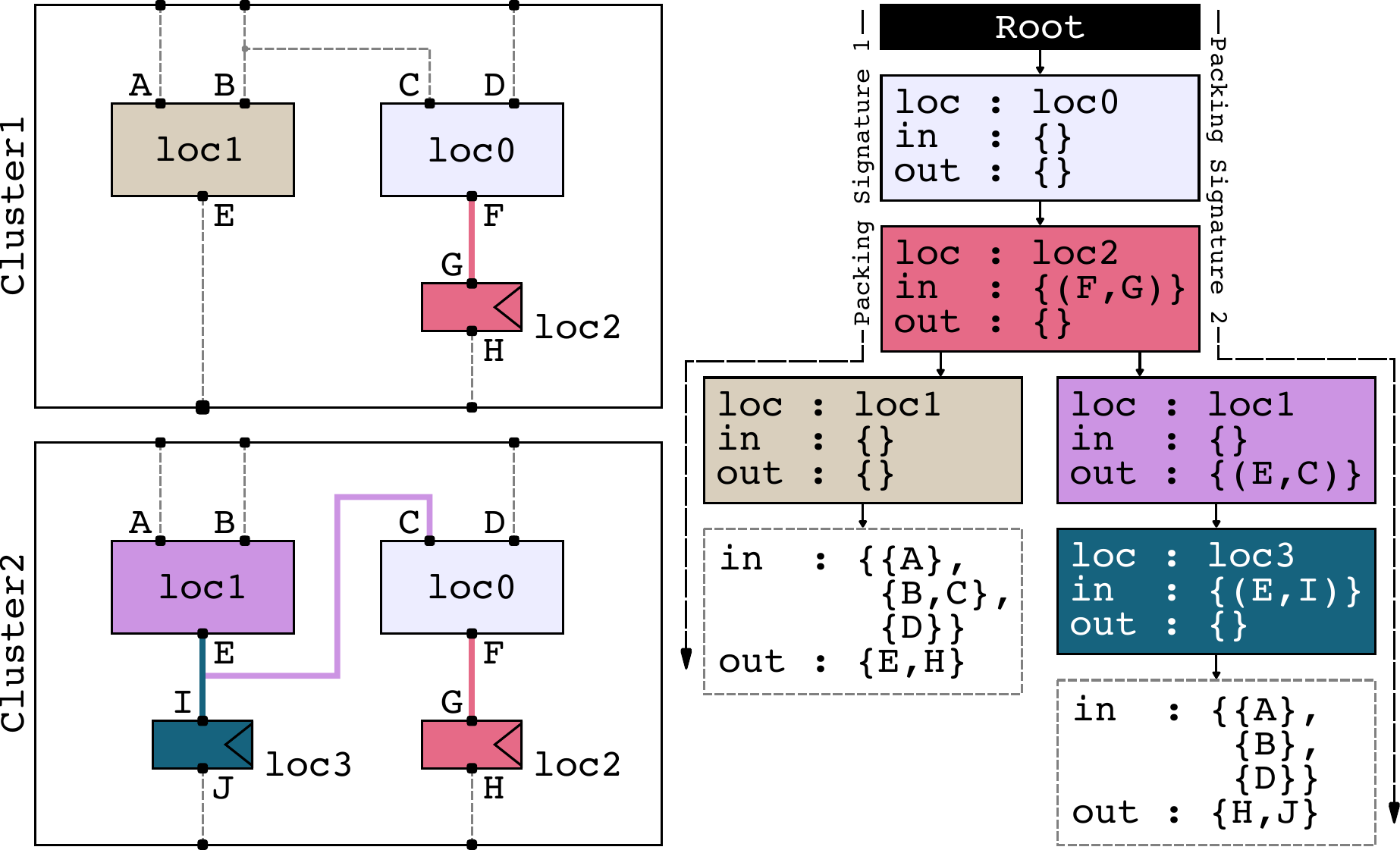}}
    \caption{Example PST representation with two packing signatures (right) corresponding to two packed clusters targeting a simplified LB architecture consisting of two 2-LUTs and two bypassable FFs (left).}
    \label{fig:pst}
\end{figure}

\begin{figure*}
    \centering{\includegraphics[width=2\columnwidth]{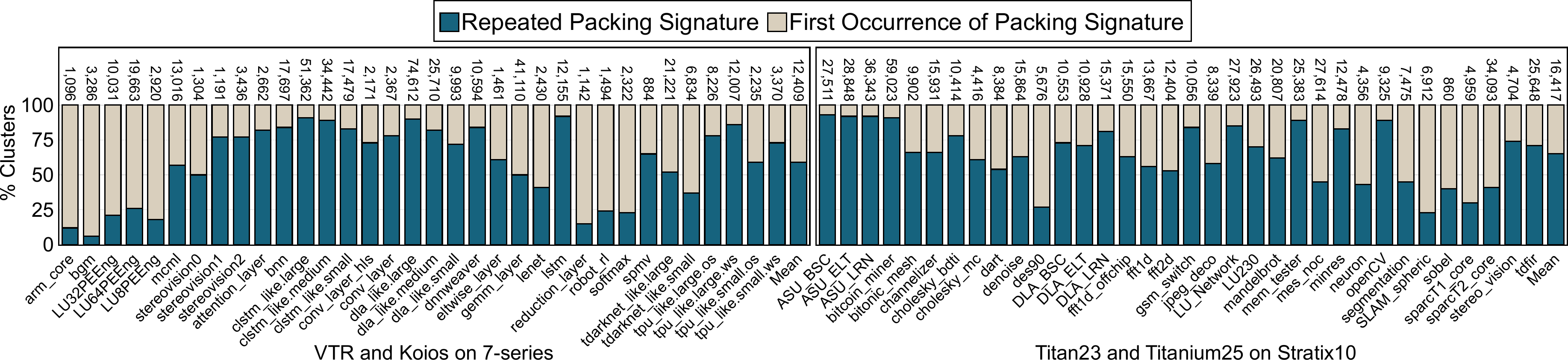}}
    \caption{Breakdown of final packed clusters that are the first occurrence vs. have a repeated packing signature. On average, 59\% and 65\% of the clusters have repeated signatures for VTR and Koios benchmarks on 7-series (left) and Titan23 and Titanium25 benchmarks on Stratix 10 (right), respectively. The total number of clusters for each benchmark is shown above its entry.}
    \label{fig:repeated-clusters}
    \vspace{-0.6cm}
\end{figure*}

The example in Fig.~\ref{fig:pst} illustrates our PST data structure in the case of a simple hypothetical LB architecture that consists of two 2-input LUTs and two bypassable FFs.
This means there are four possible locations for netlist primitives in a cluster (\texttt{loc0}-\texttt{loc3}). 
Although the primitive type that can be inserted in each location is hinted at in the figure (\texttt{loc0} and \texttt{loc1} for the LUTs, and \texttt{loc2} and \texttt{loc3} for the FFs), this information is not necessary to uniquely differentiate signatures and therefore is not included in our representation.
In addition, each pin for each of these cluster locations is given a unique identifier (\texttt{A} through \texttt{J} in the figure).
The left side of the figure depicts two different cluster candidates formed by the packing algorithm, while the right side shows how the packing signatures of these two clusters are represented in our PST.
We note that VPR represents cluster resources that support $M$ modes of operation as $M$ distinct mutually exclusive locations (e.g., a LUT-6 that can be fractured into two LUT-5 is represented as 3 locations, each with distinct pins).
Therefore, no special consideration is needed for operating modes in our representation.

Assume that \texttt{Cluster1} was the first cluster to be packed by first inserting a LUT in \texttt{loc0}, then a FF in \texttt{loc2}, and finally a LUT in \texttt{loc1}.
This sequence of packing decisions adds 3 \emph{location and connectivity nodes (LCNs)} to the PST, followed by an \emph{external connectivity node (ECN)} added to the \emph{tail} of the path labeled \texttt{Packing Signature 1} in Fig.~\ref{fig:pst}. 
When a molecule's atoms are packed into a cluster, their corresponding LCNs store the placement location as well as the internal cluster connections that drive their input pins and are driven by their output pins.

Since the packing algorithm sequentially adds molecules to a cluster, it is impossible to know at the time of packing a molecule if its connections entering/exiting the cluster will remain as external connections (i.e., that the other molecules driving or driven by it will not be packed into the same cluster later).
To avoid backtracking through the PST, only pin connections that are known to be fully contained within the cluster upon the addition of an atom are listed in its corresponding LCN; if the newly packed atom is driven by a pin belonging to a previously packed atom (e.g., \texttt{loc2} FF driven by \texttt{loc0} LUT in Fig.~\ref{fig:pst}), then a tuple with the internal source and sink pin identifiers (e.g., (\texttt{F,G})) is added to the set of input connections in the FF's LCN. 

Because the connections entering and leaving the cluster are not fully known until the cluster is finalized, their description is included in an ECN at the end of a packing signature. 
An ECN stores the internal input pins driven by external cluster inputs as a set of sets, where the union of the inner sets includes all pins that are driven by a source outside of the cluster, and pins belonging to the same inner set share the same driver. 
To illustrate this, pins \texttt{B} and \texttt{C} of \texttt{Cluster1} in Fig.~\ref{fig:pst} share the same external source, and pins \texttt{A} and \texttt{D} have independent sources. 
Therefore, the ECN at the end of the \texttt{Packing Pattern 1} branch of the PST has \texttt{B} and \texttt{C} in the same inner set. 
The ECN also stores a set of all of the internal pins that drive nets exiting the cluster.

If \texttt{Cluster2} is then packed in the following sequence: LUT in \texttt{loc0}, FF in \texttt{loc2}, LUT in \texttt{loc1}, and FF in \texttt{loc3}, the PST is traversed starting from the root as each new molecule is added to the cluster.
Since the first two steps are identical to those done during packing \texttt{Cluster1}, no new LCNs are added to the tree.
However, the third and fourth steps diverge from \texttt{Packing Pattern 1} and add two new LCNs followed by an ECN, as shown by the branch labeled \texttt{Packing Pattern 2} in Fig.~\ref{fig:pst}.
When packing a new cluster, if the PST is traversed from its root to an ECN leaf without adding any new nodes, this means that an identical cluster has been packed before. \emph{Déjà vu!}

Modern FPGA CAD tools, including VPR, will recognize that input pins to LEs and certain other primitive types are logically equivalent and, therefore, interchangeable. 
Logically, it is sufficient for packing signatures to represent equivalent inputs as a single, shared pin identifier; 
however, there are some implementation caveats with this approach. 
Heuristics based routing algorithms used by FPGA CAD tools for local and global routing, such as PathFinder~\cite{mcmurchie1995pathfinder}, can occasionally fail to route a collection of signals when a valid routing exists~\cite{shrivastava2025guaranteed}. 
Our experiments found that in rare instances, logically equivalent packing signatures could report being simultaneously legal and illegal if permutations in equivalent primitive inputs seeded the router to produce different results, and at least one routing failed. 
To address this issue, we choose not to map equivalent pins to the same pin identifier in our PST implementation in VPR. 
Instead, we simply generate LCNs using the non-random placeholder pin mappings that the technology mapper assigns to primitives with equivalent pins when constructing the netlist.

\subsection{Packing Signature Characterization}
\label{sub:pst_char}

We first implement our PST data structure in VPR and use it to enumerate the occurrences of packing signatures for the FPGA architectures and benchmarks  evaluated in Section~\ref{sec:vpr_char}.
For this experiment, we modify the PST by adding a counter variable to each ECN which gets incremented every time that ECN is visited (i.e., a legal cluster is finalized).
Fig.~\ref{fig:repeated-clusters} summarizes the results of this experiment with the total number of logic clusters for each benchmark reported on top of the bars.
It shows the ratio of unique signatures out of the final packed clusters of each circuit (i.e., count of signature first occurrences) in comparison to repeated signatures.
On average, 59\% and 65\% of the clusters packed for the evaluated benchmarks have repeated packing signatures on the 7-series and Stratix 10 architectures, respectively.
Of the tested benchmarks, approximately one-fifth (13 circuits from the Koios, Titan23, and Titanium25 suites) exhibited more than 85\% repeated clusters. 
Nine of these benchmarks come from the deep learning (DL) domain, which is known to have highly regular designs that feature large arrays of repeated compute structures.
In contrast, 53\% of clusters from non-DL Titan23, Titanium, and VTR benchmarks are repeated, on average.

%% file: 05-dejavu-packing.tex
\section{\dejavu~ Packing}
\label{sec:dejavu}

\input{results-table}

The results presented in the previous section establish three key observations.
On average across a diverse set of benchmarks targeting two VTR architectures based on modern commercial FPGAs: (1) the packing stage constitutes \mbox{58--94\%} of VPR's end-to-end runtime, depending on the architecture, (2) upwards of 90\% of this packing runtime is spent performing intracluster routing legality checks on illegal clusters, and partial clusters, during detailed packing, and (3) \mbox{59--65\%} of the final packed clusters have packing signatures that have been previously encountered during the packing process. 
The runtime impact from this last point is compounded by the numerous partially packed cluster patterns that must be evaluated for routability during intermediate steps of detailed packing (see Section~\ref{sec:vpr_packing}), many of which may have been seen before.

To eliminate redundant legality checks, we update the packer to generate and insert an ECN into the PST each time the router is invoked to assess cluster routing legality. The legality check outcome then gets stored in the ECN as an additional boolean field. Now, when the packer requires the legality status of a cluster it is packing, it will first search the child ECNs of the LCN at the tail of the active packing signature. If an ECN matches the current external connectivity state of the cluster, then the legality status recorded in that ECN is provided to the packer; otherwise, the legality status is unknown and the router must be invoked.

When an illegal cluster pattern is identified during detailed packing, the packer will remove the most recent molecule (which caused the violation) from the cluster. In turn, the pointer to the tail of the active packing signature must also be reset to its location prior to the addition of the most recent molecule.
However, the LCN(s) and ECN that were added to the PST for the violating molecule remain. This way, subsequent detailed packing attempts will find the ECN, and, seeing that is is marked as illegal, know to skip invoking the router if this pattern is reached again.

The modified packing procedure can thus be summarized as follows.
First, speculative packing is attempted as before.
As molecules are added to the cluster, the PST is traversed by following the corresponding LCNs, if they exist.
Otherwise, new LCNs are added, creating new branches in the tree.
Because speculative packing invokes routing just once when a fully packed cluster candidate has been formed, only one ECN gets generated and appended to the end of speculative packing signatures.
In the event that speculative packing fails and detailed packing is performed, an ECN is created for every molecule packed in the detailed pass and is used to store the legality result of the routing call made for the molecule.
In any state where a child ECN of the active packing signature's tail matches the external connectivity of the cluster candidate being packed, the packer can read the legality check outcome from this ECN instead of incurring the cost of performing intracluster routing.

%% file: results-table.tex
\begin{table*}[]
\centering
\footnotesize
\caption{\DejaVu~runtime gains for the packing stage and end-to-end VPR flow.}
\label{tab:packing-time}
\setlength{\tabcolsep}{0pt} 
\begin{tabular}{L{2.2cm}C{0.2cm}R{1.2cm}R{2cm}C{0.2cm}R{1.2cm}R{2cm}C{0.3cm}L{1.5cm}C{0.2cm}R{1.2cm}R{2cm}C{0.2cm}R{1.5cm}R{2.2cm}}
\cline{1-7} \cline{9-15}\\  [-1.8ex]
\multicolumn{7}{c}{\textbf{VTR \& Koios Benchmarks on 7-series}} & \textbf{} & \multicolumn{7}{c}{\textbf{Titan23 \& Titanium25 Benchmarks on Stratix 10}} \\ 

\cline{1-7} \cline{9-15}\\  [-1.8ex] 

&                          & \multicolumn{2}{c}{\textbf{Packing Runtime (mins)}}                                 & \multicolumn{1}{c}{}     & \multicolumn{2}{c}{\textbf{VPR Runtime (mins)}}                                      & \multicolumn{1}{c}{} & \multicolumn{1}{c}{}                   &                          & \multicolumn{2}{c}{\textbf{Packing Runtime (mins)}}                                   & \multicolumn{1}{c}{}     & \multicolumn{2}{c}{\textbf{VPR Runtime (mins)}}                                        \\ 


\multicolumn{1}{c}{\textbf{Benchmark}}       &                          & \multicolumn{1}{c}{\textbf{Baseline}} & \multicolumn{1}{c}{\textbf{Déjà Vu}}        & \multicolumn{1}{c}{}     & \multicolumn{1}{c}{\textbf{Baseline}} & \multicolumn{1}{c}{\textbf{Déjà Vu}}         & \multicolumn{1}{c}{} & \multicolumn{1}{c}{\textbf{Benchmark}} &                          & \multicolumn{1}{c}{\textbf{Baseline}} & \multicolumn{1}{c}{\textbf{Déjà Vu}}          & \multicolumn{1}{c}{}     & \multicolumn{1}{c}{\textbf{Baseline}} & \multicolumn{1}{c}{\textbf{Déjà Vu}}           \\ 

\cline{1-1} \cline{3-4} \cline{6-7} \cline{9-9} \cline{11-12} \cline{14-15}\\  [-1.8ex] 

arm\_core                                    &                          & 1.2                                  & 1.0 (~1.3$\times$)                          &                          & 2.3                                  & 2.0 (1.1$\times$)                           &                      & ASU\_BSC                               &                          & 2316.2                               & 79.1 (29.3$\times$)                          &                          & 2407.1                               & 196.9 (12.2$\times$)                          \\
\cellcolor[HTML]{EFEFEF}bgm                  & \cellcolor[HTML]{EFEFEF} & \cellcolor[HTML]{EFEFEF}10.7         & \cellcolor[HTML]{EFEFEF}7.2 (~1.5$\times$)  & \cellcolor[HTML]{EFEFEF} & \cellcolor[HTML]{EFEFEF}15.1         & \cellcolor[HTML]{EFEFEF}11.9 (1.3$\times$)  &                      & \cellcolor[HTML]{EFEFEF}ASU\_ELT       & \cellcolor[HTML]{EFEFEF} & \cellcolor[HTML]{EFEFEF}2460.1       & \cellcolor[HTML]{EFEFEF}87.9 (28.0$\times$)  & \cellcolor[HTML]{EFEFEF} & \cellcolor[HTML]{EFEFEF}2727.9       & \cellcolor[HTML]{EFEFEF}405.1 (~6.7$\times$)   \\
LU32PEEng                                    &                          & 41.2                                 & 21.4 (~1.9$\times$)                         &                          & 59.6                                 & 44.0 (1.4$\times$)                          &                      & ASU\_LRN                               &                          & 3782.1                               & 197.9 (19.1$\times$)                         &                          & 3839.9                               & 262.7 (14.6$\times$)                         \\
\cellcolor[HTML]{EFEFEF}LU64PEEng            & \cellcolor[HTML]{EFEFEF} & \cellcolor[HTML]{EFEFEF}69.1         & \cellcolor[HTML]{EFEFEF}36.0 (~1.9$\times$) & \cellcolor[HTML]{EFEFEF} & \cellcolor[HTML]{EFEFEF}117.4        & \cellcolor[HTML]{EFEFEF}87.2 (1.3$\times$)  &                      & \cellcolor[HTML]{EFEFEF}bitcoin\_miner & \cellcolor[HTML]{EFEFEF} & \cellcolor[HTML]{EFEFEF}4463.6       & \cellcolor[HTML]{EFEFEF}198.0 (22.5$\times$) & \cellcolor[HTML]{EFEFEF} & \cellcolor[HTML]{EFEFEF}4610.4       & \cellcolor[HTML]{EFEFEF}349.6 (13.2$\times$) \\
LU8PEEng                                     &                          & 13.1                                 & 7.5 (~1.7$\times$)                          &                          & 16.8                                 & 11.8 (1.4$\times$)                          &                      & bitonic\_mesh                          &                          & 461.9                                & 27.0 (17.1$\times$)                          &                          & 476.2                                & 46.7 (10.2$\times$)                           \\
\cellcolor[HTML]{EFEFEF}mcml                 & \cellcolor[HTML]{EFEFEF} & \cellcolor[HTML]{EFEFEF}28.9         & \cellcolor[HTML]{EFEFEF}17.8 (~1.6$\times$) & \cellcolor[HTML]{EFEFEF} & \cellcolor[HTML]{EFEFEF}65.4         & \cellcolor[HTML]{EFEFEF}59.4 (1.1$\times$)  &                      & \cellcolor[HTML]{EFEFEF}channelizer    & \cellcolor[HTML]{EFEFEF} & \cellcolor[HTML]{EFEFEF}1288.4       & \cellcolor[HTML]{EFEFEF}126.1 (10.2$\times$) & \cellcolor[HTML]{EFEFEF} & \cellcolor[HTML]{EFEFEF}1319.6       & \cellcolor[HTML]{EFEFEF}166.3 (~7.9$\times$)  \\
stereovision0                                &                          & 6.0                                 & 1.7 (~3.5$\times$)                          &                          & 7.1                                  & 2.7 (2.6$\times$)                           &                      & cholesky\_bdti                         &                          & 747.0                                & 73.0 (10.2$\times$)                          &                          & 755.6                                & 84.6 (~8.9$\times$)                            \\
\cellcolor[HTML]{EFEFEF}stereovision1        & \cellcolor[HTML]{EFEFEF} & \cellcolor[HTML]{EFEFEF}5.5          & \cellcolor[HTML]{EFEFEF}0.4 (13.0$\times$) & \cellcolor[HTML]{EFEFEF} & \cellcolor[HTML]{EFEFEF}7.8          & \cellcolor[HTML]{EFEFEF}2.6 (3.0$\times$)   &                      & \cellcolor[HTML]{EFEFEF}cholesky\_mc   & \cellcolor[HTML]{EFEFEF} & \cellcolor[HTML]{EFEFEF}325.2        & \cellcolor[HTML]{EFEFEF}61.3 (~5.3$\times$)   & \cellcolor[HTML]{EFEFEF} & \cellcolor[HTML]{EFEFEF}329.7        & \cellcolor[HTML]{EFEFEF}66.1 (~5.0$\times$)    \\
stereovision2                                &                          & 16.5                                 & 2.0 (~8.3$\times$)                          &                          & 22.5                                 & 9.8 (2.3$\times$)                           &                      & dart                                   &                          & 154.6                                & 43.0 (~3.6$\times$)                           &                          & 163.4                                & 53.3 (~3.1$\times$)                            \\
\cellcolor[HTML]{EFEFEF}attention\_layer     & \cellcolor[HTML]{EFEFEF} & \cellcolor[HTML]{EFEFEF}14.6         & \cellcolor[HTML]{EFEFEF}1.8 (~8.0$\times$)  & \cellcolor[HTML]{EFEFEF} & \cellcolor[HTML]{EFEFEF}39.3         & \cellcolor[HTML]{EFEFEF}30.7 (1.3$\times$)  &                      & \cellcolor[HTML]{EFEFEF}denoise        & \cellcolor[HTML]{EFEFEF} & \cellcolor[HTML]{EFEFEF}169.4        & \cellcolor[HTML]{EFEFEF}60.7 (~2.8$\times$)   & \cellcolor[HTML]{EFEFEF} & \cellcolor[HTML]{EFEFEF}215.6        & \cellcolor[HTML]{EFEFEF}110.1 (~2.0$\times$)   \\
bnn                                          &                          & 24.7                                 & 5.3 (~4.7$\times$)                          &                          & 59.9                                 & 48.4 (1.2$\times$)                          &                      & des90                                  &                          & 392.5                                & 67.0 (~5.9$\times$)                           &                          & 399.7                                & 74.9 (~5.3$\times$)                            \\
\cellcolor[HTML]{EFEFEF}clstm\_like.large    & \cellcolor[HTML]{EFEFEF} & \cellcolor[HTML]{EFEFEF}81.3         & \cellcolor[HTML]{EFEFEF}13.8 (~5.9$\times$) & \cellcolor[HTML]{EFEFEF} & \cellcolor[HTML]{EFEFEF}211.8        & \cellcolor[HTML]{EFEFEF}154.7 (1.4$\times$) &                      & \cellcolor[HTML]{EFEFEF}DLA\_BSC       & \cellcolor[HTML]{EFEFEF} & \cellcolor[HTML]{EFEFEF}1122.1       & \cellcolor[HTML]{EFEFEF}129.1 (~8.7$\times$)  & \cellcolor[HTML]{EFEFEF} & \cellcolor[HTML]{EFEFEF}1131.3       & \cellcolor[HTML]{EFEFEF}141.4 (~8.0$\times$)   \\
clstm\_like.medium                           &                          & 55.9                                 & 9.7 (~5.8$\times$)                          &                          & 131.4                                & 93.9 (1.4$\times$)                          &                      & DLA\_ELT                               &                          & 1124.1                              & 140.6 (~8.0$\times$)                          &                          & 1133.5                               & 152.8 (~7.4$\times$)                           \\
\cellcolor[HTML]{EFEFEF}clstm\_like.small    & \cellcolor[HTML]{EFEFEF} & \cellcolor[HTML]{EFEFEF}31.3         & \cellcolor[HTML]{EFEFEF}6.0 (~5.2$\times$)  & \cellcolor[HTML]{EFEFEF} & \cellcolor[HTML]{EFEFEF}60.3         & \cellcolor[HTML]{EFEFEF}42.0 (1.4$\times$)  &                      & \cellcolor[HTML]{EFEFEF}DLA\_LRN       & \cellcolor[HTML]{EFEFEF} & \cellcolor[HTML]{EFEFEF}1858.6       & \cellcolor[HTML]{EFEFEF}140.5 (13.2$\times$) & \cellcolor[HTML]{EFEFEF} & \cellcolor[HTML]{EFEFEF}1872.0       & \cellcolor[HTML]{EFEFEF}159.9 (11.7$\times$)  \\
conv\_layer\_hls                             &                          & 2.4                                  & 1.3 (~1.9$\times$)                          &                          & 8.4                                  & 7.4 (1.1$\times$)                           &                      & fft1d\_offchip                         &                          & 1213.2                               & 141.5 (~8.6$\times$)                          &                          & 1243.1                               & 178.7 (~7.0$\times$)                           \\
\cellcolor[HTML]{EFEFEF}conv\_layer          & \cellcolor[HTML]{EFEFEF} & \cellcolor[HTML]{EFEFEF}4.5          & \cellcolor[HTML]{EFEFEF}0.8 (~5.6$\times$)  & \cellcolor[HTML]{EFEFEF} & \cellcolor[HTML]{EFEFEF}7.9          & \cellcolor[HTML]{EFEFEF}4.1 (1.9$\times$)   &                      & \cellcolor[HTML]{EFEFEF}fft1d          & \cellcolor[HTML]{EFEFEF} & \cellcolor[HTML]{EFEFEF}1182.6       & \cellcolor[HTML]{EFEFEF}122.4 (~9.7$\times$)  & \cellcolor[HTML]{EFEFEF} & \cellcolor[HTML]{EFEFEF}1199.2       & \cellcolor[HTML]{EFEFEF}144.9 (~8.3$\times$)   \\
dla\_like.large                              &                          & 169.5                                & 29.6 (~5.7$\times$)                         &                          & 462.0                                & 345.5 (1.3$\times$)                         &                      & fft2d                                  &                          & 1010.7                               & 113.7 (~8.9$\times$)                          &                          & 1028.6                               & 137.8 (~7.5$\times$)                           \\
\cellcolor[HTML]{EFEFEF}dla\_like.medium     & \cellcolor[HTML]{EFEFEF} & \cellcolor[HTML]{EFEFEF}86.2         & \cellcolor[HTML]{EFEFEF}15.7 (~5.5$\times$) & \cellcolor[HTML]{EFEFEF} & \cellcolor[HTML]{EFEFEF}152.6        & \cellcolor[HTML]{EFEFEF}95.5 (1.6$\times$)  &                      & \cellcolor[HTML]{EFEFEF}gsm\_switch    & \cellcolor[HTML]{EFEFEF} & \cellcolor[HTML]{EFEFEF}134.7        & \cellcolor[HTML]{EFEFEF}24.8 (~5.4$\times$)   & \cellcolor[HTML]{EFEFEF} & \cellcolor[HTML]{EFEFEF}150.9        & \cellcolor[HTML]{EFEFEF}43.4 (~3.5$\times$)    \\
dla\_like.small                              &                          & 52.7                                 & 10.3 (~5.1$\times$)                         &                          & 70.3                                 & 35.5 (2.0$\times$)                          &                      & jpeg\_deco                             &                          & 424.5                                & 105.4 (~4.0$\times$)                          &                          & 443.6                                & 125.7 (~3.5$\times$)                           \\
\cellcolor[HTML]{EFEFEF}dnnweaver            & \cellcolor[HTML]{EFEFEF} & \cellcolor[HTML]{EFEFEF}20.8         & \cellcolor[HTML]{EFEFEF}7.3 (~2.8$\times$)  & \cellcolor[HTML]{EFEFEF} & \cellcolor[HTML]{EFEFEF}51.0         & \cellcolor[HTML]{EFEFEF}42.7 (1.2$\times$)  &                      & \cellcolor[HTML]{EFEFEF}LU\_Network    & \cellcolor[HTML]{EFEFEF} & \cellcolor[HTML]{EFEFEF}1046.9       & \cellcolor[HTML]{EFEFEF}89.3 (11.7$\times$)  & \cellcolor[HTML]{EFEFEF} & \cellcolor[HTML]{EFEFEF}1085.1       & \cellcolor[HTML]{EFEFEF}138.3 (~7.8$\times$)   \\
eltwise\_layer                               &                          & 7.2                                 & 2.2 (~3.3$\times$)                          &                          & 9.51                                  & 4.6 (2.1$\times$)                           &                      & LU230                                  &                          & 845.0                                & 131.3 (~6.4$\times$)                          &                          & 882.2                                & 179.3 (~4.9$\times$)                           \\
\cellcolor[HTML]{EFEFEF}gemm\_layer          & \cellcolor[HTML]{EFEFEF} & \cellcolor[HTML]{EFEFEF}103.1        & \cellcolor[HTML]{EFEFEF}31.5 (~3.3$\times$) & \cellcolor[HTML]{EFEFEF} & \cellcolor[HTML]{EFEFEF}191.5        & \cellcolor[HTML]{EFEFEF}130.8 (1.5$\times$) &                      & \cellcolor[HTML]{EFEFEF}mandelbrot     & \cellcolor[HTML]{EFEFEF} & \cellcolor[HTML]{EFEFEF}1828.4       & \cellcolor[HTML]{EFEFEF}191.8 (~9.5$\times$)  & \cellcolor[HTML]{EFEFEF} & \cellcolor[HTML]{EFEFEF}1851.3       & \cellcolor[HTML]{EFEFEF}220.6 (~8.4$\times$)   \\
lenet                                        &                          & 8.1                                  & 3.6 (~2.2$\times$)                          &                          & 10.9                                 & 6.7 (1.6$\times$)                           &                      & mem\_tester                            &                          & 527.2                                & 42.0 (12.6$\times$)                          &                          & 550.8                                & 72.2 (~7.6$\times$)                            \\
\cellcolor[HTML]{EFEFEF}lstm                 & \cellcolor[HTML]{EFEFEF} & \cellcolor[HTML]{EFEFEF}80.1         & \cellcolor[HTML]{EFEFEF}6.8 (11.9$\times$) & \cellcolor[HTML]{EFEFEF} & \cellcolor[HTML]{EFEFEF}106.0        & \cellcolor[HTML]{EFEFEF}40.0 (2.7$\times$)  &                      & \cellcolor[HTML]{EFEFEF}mes\_noc       & \cellcolor[HTML]{EFEFEF} & \cellcolor[HTML]{EFEFEF}421.4        & \cellcolor[HTML]{EFEFEF}118.8 (~3.5$\times$)  & \cellcolor[HTML]{EFEFEF} & \cellcolor[HTML]{EFEFEF}466.5        & \cellcolor[HTML]{EFEFEF}169.1 (~2.8$\times$)   \\
reduction\_layer                             &                          & 8.5                                  & 4.6 (~1.8$\times$)                          &                          & 9.8                                  & 6.0 (1.6$\times$)                           &                      & minres                                 &                          & 680.3                                & 58.3 (11.7$\times$)                          &                          & 689.6                                & 70.6 (~9.8$\times$)                            \\
\cellcolor[HTML]{EFEFEF}robot\_rl            & \cellcolor[HTML]{EFEFEF} & \cellcolor[HTML]{EFEFEF}3.9          & \cellcolor[HTML]{EFEFEF}0.8 (~4.6$\times$)  & \cellcolor[HTML]{EFEFEF} & \cellcolor[HTML]{EFEFEF}5.6          & \cellcolor[HTML]{EFEFEF}2.6 (2.2$\times$)   &                      & \cellcolor[HTML]{EFEFEF}neuron         & \cellcolor[HTML]{EFEFEF} & \cellcolor[HTML]{EFEFEF}380.1        & \cellcolor[HTML]{EFEFEF}97.3 (~3.9$\times$)   & \cellcolor[HTML]{EFEFEF} & \cellcolor[HTML]{EFEFEF}383.1        & \cellcolor[HTML]{EFEFEF}100.4 (~3.8$\times$)   \\
softmax                                      &                          & 9.8                                  & 5.0 (~1.9$\times$)                          &                          & 12.1                                 & 7.7 (1.6$\times$)                           &                      & openCV                                 &                          & 383.0                                & 29.5 (13.0$\times$)                          &                          & 395.1                                & 43.6 (~9.1$\times$)                            \\
\cellcolor[HTML]{EFEFEF}spmv                 & \cellcolor[HTML]{EFEFEF} & \cellcolor[HTML]{EFEFEF}3.9          & \cellcolor[HTML]{EFEFEF}0.9 (~4.2$\times$)  & \cellcolor[HTML]{EFEFEF} & \cellcolor[HTML]{EFEFEF}9.6          & \cellcolor[HTML]{EFEFEF}6.8 (1.4$\times$)   &                      & \cellcolor[HTML]{EFEFEF}segmentation   & \cellcolor[HTML]{EFEFEF} & \cellcolor[HTML]{EFEFEF}79.1         & \cellcolor[HTML]{EFEFEF}42.5 (~1.9$\times$)   & \cellcolor[HTML]{EFEFEF} & \cellcolor[HTML]{EFEFEF}93.4         & \cellcolor[HTML]{EFEFEF}58.1 (~1.6$\times$)    \\
tdarknet\_like.large                         &                          & 66.9                                 & 24.4 (~2.7$\times$)                         &                          & 133.0                                & 95.2 (1.4$\times$)                          &                      & SLAM\_spheric                          &                          & 29.0                                 & 23.0 (~1.3$\times$)                           &                          & 42.0                                 & 37.5 (~1.1$\times$)                            \\
\cellcolor[HTML]{EFEFEF}tdarknet\_like.small & \cellcolor[HTML]{EFEFEF} & \cellcolor[HTML]{EFEFEF}27.1         & \cellcolor[HTML]{EFEFEF}12.5 (~2.2$\times$) & \cellcolor[HTML]{EFEFEF} & \cellcolor[HTML]{EFEFEF}75.4         & \cellcolor[HTML]{EFEFEF}67.4 (1.1$\times$)  &                      & \cellcolor[HTML]{EFEFEF}sobel          & \cellcolor[HTML]{EFEFEF} & \cellcolor[HTML]{EFEFEF}29.2         & \cellcolor[HTML]{EFEFEF}15.0 (~1.9$\times$)   & \cellcolor[HTML]{EFEFEF} & \cellcolor[HTML]{EFEFEF}32.6         & \cellcolor[HTML]{EFEFEF}18.4 (~1.8$\times$)    \\
tpu\_like.large.os                           &                          & 69.6                                 & 5.2 (13.4$\times$)                         &                          & 96.0                                 & 39.1 (2.5$\times$)                          &                      & sparcT1\_core                          &                          & 55.5                                 & 43.0 (~1.3$\times$)                           &                          & 62.3                                 & 50.3 (~1.2$\times$)                            \\
\cellcolor[HTML]{EFEFEF}tpu\_like.large.ws   & \cellcolor[HTML]{EFEFEF} & \cellcolor[HTML]{EFEFEF}9.6          & \cellcolor[HTML]{EFEFEF}3.0 (~3.2$\times$)  & \cellcolor[HTML]{EFEFEF} & \cellcolor[HTML]{EFEFEF}40.8         & \cellcolor[HTML]{EFEFEF}38.2 (1.1$\times$)  &                      & \cellcolor[HTML]{EFEFEF}sparcT2\_core  & \cellcolor[HTML]{EFEFEF} & \cellcolor[HTML]{EFEFEF}355.6        & \cellcolor[HTML]{EFEFEF}211.3 (~1.7$\times$)  & \cellcolor[HTML]{EFEFEF} & \cellcolor[HTML]{EFEFEF}407.5        & \cellcolor[HTML]{EFEFEF}262.0 (~1.6$\times$)   \\
tpu\_like.small.os                           &                          & 20.6                                 & 2.5 (~8.1$\times$)                          &                          & 26.9                                 & 10.6 (2.5$\times$)                          &                      & stereo\_vision                         &                          & 474.6                                & 69.5 (~6.8$\times$)                           &                          & 481.9                                & 78.2 (~6.2$\times$)                            \\
\cellcolor[HTML]{EFEFEF}tpu\_like.small.ws   & \cellcolor[HTML]{EFEFEF} & \cellcolor[HTML]{EFEFEF}5.8          & \cellcolor[HTML]{EFEFEF}1.8 (~3.2$\times$)  & \cellcolor[HTML]{EFEFEF} & \cellcolor[HTML]{EFEFEF}13.7         & \cellcolor[HTML]{EFEFEF}10.3 (1.3$\times$)  &                      & \cellcolor[HTML]{EFEFEF}tdfir          & \cellcolor[HTML]{EFEFEF} & \cellcolor[HTML]{EFEFEF}1278.7       & \cellcolor[HTML]{EFEFEF}130.6 (~9.8$\times$)  & \cellcolor[HTML]{EFEFEF} & \cellcolor[HTML]{EFEFEF}1310.7       & \cellcolor[HTML]{EFEFEF}171.2 (~7.7$\times$)   \\ \cline{1-7} \cline{9-15} \\  [-1.8ex]
\textbf{Geomean}                             &                          & \multicolumn{1}{l}{}                  & \textbf{(3.75$\times$)}                     & \multicolumn{1}{l}{}     & \multicolumn{1}{l}{}                  & \textbf{(1.57$\times$)}                      & \textbf{}            & \textbf{Geomean}                       &                          & \multicolumn{1}{l}{}                  & \textbf{(6.92$\times$)}                       & \multicolumn{1}{l}{}     & \multicolumn{1}{l}{}                  & \textbf{(5.31$\times$)}                        \\ \cline{1-7} \cline{9-15} \\  [-1.8ex]
\end{tabular}
\vspace{-0.3cm}
\end{table*}

%% file: 06-results.tex
\section{Experimental Results}
\label{sec:results}

\begin{figure*}
    \centering{\includegraphics[width=2\columnwidth]{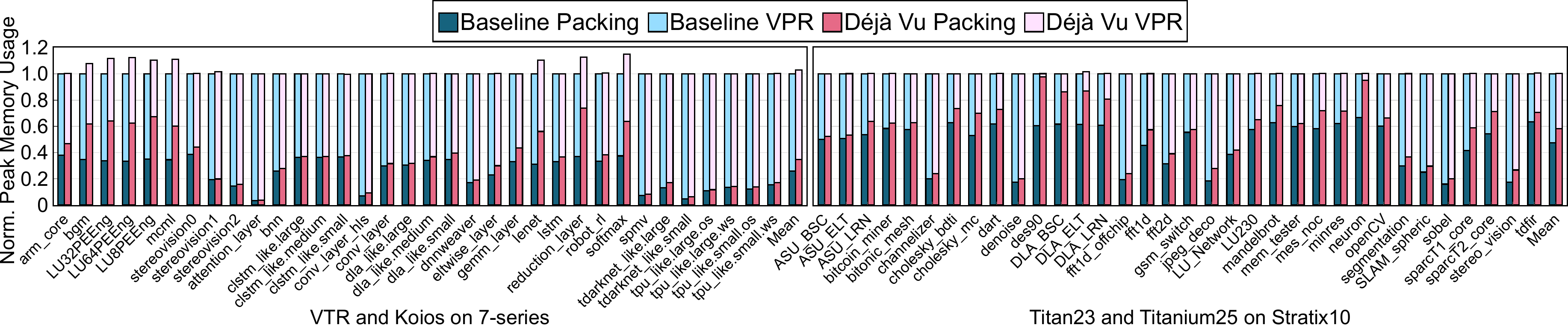}}
    \caption{Comparison of peak memory utilization of the VPR packing stage and full VPR flow with and without \dejavu~packing. All values are normalized to the baseline peak VPR memory utilization. Packing memory utilization increased by 30\% and 24\% on average for the VTR and Koios benchmarks on 7-series (left) and Titan23 and Titanium25 benchmarks on Stratix 10 (right). VPR peak memory utilization increased by 3\% and <1\% on average.}
    \label{fig:memory_usage}
    \vspace{-0.5cm}
\end{figure*}

Table \ref{tab:packing-time} shows the runtime gains achieved by \DejaVu~packing for the VTR and Koios benchmarks on the 7-series architecture, as well as for the Titan23 and Titanium25 benchmarks on Stratix 10. 
Our \DejaVu~packing technique results in significant runtime improvements across every benchmark, while preserving identical QoR to the baseline VPR since the packing solution is identical.

The 7-series benchmarks see speedups of up to 13.4$\times$ in the packing stage and 3.0$\times$ for the entire VPR flow. 
For the two Koios benchmarks with the longest runtimes, \texttt{dla\_like.large} and \texttt{clstm\_like.large}, the packing stage is 5.7$\times$ and 5.9$\times$ faster, resulting in VPR runtime reductions of 1.3$\times$ and 1.4$\times$, respectively.
To put this in perspective, the total runtime of of \texttt{dla\_like.large} is reduced by 1.9 hours, bringing the longest runtime across all 7-series benchmarks from 7.7 hours down to 5.8 hours. 
On average, total VPR runtime is reduced by 1.6$\times$.

Our baseline runtime measurements for the Stratix 10 capture show that modeling modern commercial LB architectures in VTR results in a significant overhead in packing runtime.
On average across the Titan23 and Titanium25 benchmarks, packing takes 16.2 hours, with the largest circuit, \texttt{bitcoin\_miner}, spending 74.4 hours out of the 76.8-hour end-to-end runtime in the packing stage. 
As a result, a faithful Stratix 10 capture becomes impractical to use for architecture and CAD algorithm exploration.
The introduction of \DejaVu~packing yields runtime reductions of up to 29.3$\times$ in the packing stage.
This brings the maximum VPR runtime across all Stratix 10 benchmarks down from 76.8 hours to around 6.8 hours, and the average total VPR runtime to 2.2 hours down from 16.8 hours, well within acceptable limits for mapping large circuits in modern FPGA CAD tools.
On average, for the Stratix 10 architecture, \textbf{\DejaVu~packing speeds up the entire VPR flow by 5.3$\times$}.

Managing the PST structure in the packing stage introduces very little overhead. We measured that insertion and lookup into the PST accounts for only 0.8\% of total packing runtime on average (with a maximum of 1.6\%) across all of the evaluated 7-series and Stratix 10 benchmarks.
Fig. \ref{fig:memory_usage} compares the peak memory usage for the packing stage and the full VPR flow with and without our PST data structure. 
On average, the inclusion of a PST for \DejaVu~packing results in a 27\% increase in peak memory utilization for the packer. 
From the measured memory deltas between the baseline and \DejaVu~packing stages, we estimate that PST sizes range from \mbox{19--1258} MB for the 7-series benchmarks, and {\mbox{21--657} MB for the Stratix 10 benchmarks. 
The PST structure is freed at the end of the packing stage, prior to later VPR stages which use greater peak memory, thus the peak VPR memory usage is generally unaffected by \DejaVu~packing. 
Having verified the absence of memory leaks with Valgrind~\cite{valgrind}, the peak VPR memory increases that can be seen in a number of the 7-series benchmarks are best attributed to underlying memory allocator behavior, and not persisting PST structures.

\DejaVu~packing is most useful for architectures with complex LB structures that result in a significant amount of legality checks during detailed packing.
However, we also evaluated its use on the VTR flagship architecture that has a simple LB with a full local crossbar.
In such architectures, the overwhelming majority of their packed clusters succeed with speculative packing, since any candidate packed cluster is legal if it uses no more than the available LB pins.
We confirm that \DejaVu~packing does not introduce any runtime overhead for these architectures;
the packing stage is 3\% faster on average across the VTR and Koios benchmarks on the flagship architecture, which does not result in any noticeable change in the end-to-end runtime.
In addition, since the memory footprint of the PST scales primarily with failed legalization checks (due to the ECNs generated by detailed packing), an average of only 0.8\% increase in peak packing memory utilization (with a maximum of 2.6\%) is measured for the VTR flagship architecture.
This is demonstrates the desirable property that the memory overhead of \DejaVu~packing is tightly coupled to the runtime savings that it can achieve.

%% file: 07-conclusion.tex
\section{Conclusion}
\label{sec:conc}

We presented \DejaVu~packing, a runtime optimization for FPGA packing algorithms, particularly for architectures with complex logic block (LB) internals that resemble contemporary commercial devices.
Through detailed runtime analysis of the VPR tool, we demonstrated that packing dominates the overall flow runtime for such architectures, and that this runtime is predominantly spent performing legality checks that evaluate the routing feasibility of candidate clusters.
Our key insight is that, on average, 62\% of the final packed clusters in the benchmarks we evaluated are recurring packing patterns, each of which can invoke numerous costly legality checks on intermediate clustering steps.
We introduced a novel packing signature tree (PST) data structure, which enables efficient memoization and comparison of cluster packing patterns.
By integrating the PST into VPR's packing algorithm, we enable the packer to recognize packing patterns that have been seen before (i.e., \dejavu~instances) and skip their redundant legality checks by reusing previously computed outcomes.
We evaluated our approach across 68 benchmarks from the VTR, Koios, Titan23, and Titanium suites, targeting AMD 7-series and Altera Stratix 10 VTR architecture captures.
Our results demonstrate substantial runtime improvements: up to 29.3$\times$ reduction in packing time, with average speedups of 5.1$\times$ for packing and 2.9$\times$ for the end-to-end VPR flow.
These runtime gains are achieved while fully preserving the quality of results, as our approach does not alter the optimization decisions of the packing algorithm.
This work is being integrated into the VPR open-source tool to accelerate architecture and CAD algorithm exploration for modern FPGAs with complex LB structures.

%% file: references.bib
@article{elgammal2025vtr9,
  title={{VTR 9: Open-source CAD for Fabric and Beyond FPGA Architecture Exploration}},
  author={Elgammal, Mohamed A and Mohaghegh, Amin and Shahrouz, Soheil Gholami and Mahmoudi, Fatemehsadat and Ko{\c{s}}ar, Fahrican and Talaei, Kimia and Fife, Joshua and Khadivi, Daniel and Murray, Kevin and Boutros, Andrew and others},
  journal={ACM Transactions on Reconfigurable Technology and Systems},
  volume={18},
  number={3},
  pages={1--53},
  year={2025},
  publisher={ACM New York, NY}
}

@article{Luu2011AAPack,
  title={{Architecture description and packing for logic blocks with hierarchy, modes and complex interconnect}},
  author={Luu, Jason and Anderson, Jason and Rose, Jonathan},
  journal={Proceedings of 2011 Acm/Sigda International Symposium on Field Programmable Gate Arrays},
  pages={227-236},
  year={2011},
  publisher={ACM New York, NY}
}

@article{Luu2014vtr7,
  title={{VTR 7.0: Next Generation Architecture and CAD System for FPGAs}},
  author = {Luu, Jason and Goeders, Jeffrey and Wainberg, Michael and Somerville, Andrew and Yu, Thien and Nasartschuk, Konstantin and Nasr, Miad and Wang, Sen and Liu, Tim and Ahmed, Nooruddin and Kent, Kenneth B. and Anderson, Jason and Rose, Jonathan and Betz, Vaughn},
  journal={ACM Transactions on Reconfigurable Technology and Systems},
  volume={7},
  number={3},
  pages={1-30},
  year={2014},
  publisher={ACM New York, NY}
}

@article{murray2020vtr8,
  title={{VTR 8: High-performance CAD and Customizable FPGA Architecture Modelling}},
  author = {Murray, Kevin E. and Petelin, Oleg and Zhong, Sheng and Wang, Jia Min and Eldafrawy, Mohamed and Legault, Jean-Philippe and Sha, Eugene and Graham, Aaron G. and Wu, Jean and Walker, Matthew J. P. and Zeng, Hanqing and Patros, Panagiotis and Luu, Jason and Kent, Kenneth B. and Betz, Vaughn},
  journal={ACM Transactions on Reconfigurable Technology and Systems},
  volume={13},
  number={2},
  pages={1-60},
  year={2020},
  publisher={ACM New York, NY}
}

@inproceedings{betz1997cluster,
  title={{Cluster-based Logic Blocks for FPGAs: Area-Efficiency vs. Input Sharing and Size}},
  author={Betz, Vaughn and Rose, Jonathan},
  booktitle={IEEE Custom Integrated Circuits Conference (CICC)},
  pages={551--554},
  year={1997}
}

@inproceedings{marquardt1999using,
  title={{Using Cluster-Based Logic Blocks and Timing-Driven Packing to Improve FPGA Speed and Density}},
  author={Marquardt, Alexander and Betz, Vaughn and Rose, Jonathan},
  booktitle={ACM/SIGDA International Symposium on Field-Programmable Gate Arrays (FPGA)},
  pages={37--46},
  year={1999}
}

@inproceedings{bozorgzadeh2001rpack,
  title={{RPack: Routability-Driven Packing for Cluster-Based FPGAs}},
  author={Bozorgzadeh, Elaheh and Ogrenci-Memik, Seda and Sarrafzadeh, Majid},
  booktitle={ACM/IEEE Asia and South Pacific Design Automation Conference (ASP-DAC)},
  pages={629--634},
  year={2001}
}

@article{singh2002efficient,
  title={{Efficient Circuit Clustering for Area and Power Reduction in FPGAs}},
  author={Singh, Amit and Parthasarathy, Ganapathy and Marek-Sadowska, Malgorzata},
  journal={ACM Transactions on Design Automation of Electronic Systems (TODAES)},
  volume={7},
  number={4},
  pages={643--663},
  year={2002},
  publisher={ACM New York, NY, USA}
}

@inproceedings{chen2007improving,
  title={{Improving Timing-Driven FPGA Packing with Physical Information}},
  author={Chen, Doris T and Vorwerk, Kristofer and Kennings, Andrew},
  booktitle={IEEE International Conference on Field Programmable Logic and Applications (FPL)},
  pages={117--123},
  year={2007}
}

@inproceedings{rajavel2011mo,
  title={{MO-Pack: Many-Objective Clustering for FPGA CAD}},
  author={Rajavel, Senthilkumar Thoravi and Akoglu, Ali},
  booktitle={ACM/IEEE Design Automation Conference (DAC)},
  pages={818--823},
  year={2011}
}

@inproceedings{tom2006dopack,
  title={{Un/DoPack: Re-clustering of Large System-on-Chip Designs with Interconnect Variation for Low-Cost FPGAs}},
  author={Tom, Marvin and Leong, David and Lemieux, Guy},
  booktitle={IEEE/ACM International Conference on Computer-aided design (ICCAD)},
  pages={680--687},
  year={2006}
}

@inproceedings{liu2009t,
  title={{T-NDPack: Timing-Driven Non-Uniform Depopulation Based Clustering}},
  author={Liu, Hanyu and Akoglu, Ali},
  booktitle={IEEE Southern Conference on Programmable Logic (SPL)},
  pages={9--14},
  year={2009}
}

@inproceedings{vercruyce2016runtime,
  title={{Runtime-Quality Tradeoff in Partitioning Based Multithreaded Packing}},
  author={Vercruyce, Dries and Vansteenkiste, Elias and Stroobandt, Dirk},
  booktitle={IEEE International Conference on Field Programmable Logic and Applications (FPL)},
  pages={1--9},
  year={2016}
}

@article{vercruyce2017preserving,
  title={{How Preserving Circuit Design Hierarchy During FPGA Packing Leads to Better Performance}},
  author={Vercruyce, Dries and Vansteenkiste, Elias and Stroobandt, Dirk},
  journal={IEEE Transactions on Computer-Aided Design of Integrated Circuits and Systems (TCAD)},
  volume={37},
  number={3},
  pages={629--642},
  year={2017},
  publisher={IEEE}
}

@inproceedings{chen2004simultaneous,
  title={{Simultaneous Timing Driven Clustering and Placement for FPGAs}},
  author={Chen, Gang and Cong, Jason},
  booktitle={International Conference on Field Programmable Logic and Applications (FPL)},
  pages={158--167},
  year={2004}
}

@inproceedings{feng2012k,
  title={{K-Way Partitioning Based Packing for FPGA Logic Blocks Without Input Bandwidth Constraint}},
  author={Feng, Wenyi},
  booktitle={IEEE International Conference on Field-Programmable Technology (FPT)},
  pages={8--15},
  year={2012}
}

@inproceedings{feng2014rent,
  title={{Rent's Rule Based FPGA Packing for Routability Optimization}},
  author={Feng, Wenyi and Greene, Jonathan and Vorwerk, Kristofer and Pevzner, Val and Kundu, Arun},
  booktitle={ACM/SIGDA International Symposium on Field-Programmable Gate Arrays (FPGA)},
  pages={31--34},
  year={2014}
}

@inproceedings{elgammal2023breaking,
  title={{Breaking Boundaries: Optimizing FPGA CAD with Flexible and Multi-threaded Re-Clustering}},
  author={Elgammal, Mohamed A and Betz, Vaughn},
  booktitle={ACM International Symposium on Highly Efficient Accelerators and Reconfigurable Technologies (HEART)},
  pages={11--18},
  year={2023}
}

@inproceedings{singhal2017lsc,
  title={{LSC: A Large-Scale Consensus-Based Clustering Algorithm for High-Performance FPGAs}},
  author={Singhal, Love and Iyer, Mahesh A and Adya, Saurabh},
  booktitle={ACM/IEEE Design Automation Conference (DAC)},
  pages={1--6},
  year={2017}
}

@inproceedings{mcmurchie1995pathfinder,
    title = {{PathFinder: A Negotiation-Based Performance-Driven Router for FPGAs}},
    author = {McMurchie, Larry and Ebeling, Carl},
    booktitle = {ACM International Symposium on Field-Programmable Gate Arrays (FPGA)},
    pages = {111–-117},
    year = {1995}
}

@inproceedings{khoozani2023titan,
  title={{Titan 2.0: Enabling Open-Source CAD Evaluation with a Modern Architecture Capture}},
  author={Khoozani, Kimia Talaei and Dehkordi, Arash Ahmadian and Betz, Vaughn},
  booktitle={IEEE International Conference on Field-Programmable Logic and Applications (FPL)},
  pages={57--64},
  year={2023}
}

@inproceedings{murray2013titan,
  title={{Titan: Enabling Large and Complex Benchmarks in Academic CAD}},
  author={Murray, Kevin E and Whitty, Scott and Liu, Suya and Luu, Jason and Betz, Vaughn},
  booktitle={IEEE International Conference on Field programmable Logic and Applications (FPL)},
  pages={1--8},
  year={2013}
}

@article{arora2023koios,
  title={{Koios 2.0: Open-Source Deep Learning Benchmarks for FPGA Architecture and CAD Research}},
  author={Arora, Aman and Boutros, Andrew and Damghani, Seyed Alireza and Mathur, Karan and Mohanty, Vedant and Anand, Tanmay and Elgammal, Mohamed A and Kent, Kenneth B and Betz, Vaughn and John, Lizy K},
  journal={IEEE Transactions on Computer-Aided Design of Integrated Circuits and Systems (TCAD)},
  volume={42},
  number={11},
  pages={3895--3909},
  year={2023},
  publisher={IEEE}
}

@inproceedings{karypis1999multilevel,
  title={{Multilevel K-Way Hypergraph Partitioning}},
  author={Karypis, George and Kumar, Vipin},
  booktitle={ACM/IEEE Design Automation Conference (DAC)},
  pages={343--348},
  year={1999}
}

@article{murray2020optimizing,
  title={{Optimizing FPGA Logic Block Architectures for Arithmetic}},
  author={Murray, Kevin E and Luu, Jason and Walker, Matthew JP and McCullough, Conor and Wang, Sen and Huda, Safeen and Yan, Bo and Chiasson, Charles and Kent, Kenneth B and Anderson, Jason and others},
  journal={IEEE Transactions on Very Large Scale Integration (VLSI) Systems},
  volume={28},
  number={6},
  pages={1378--1391},
  year={2020},
  publisher={IEEE}
}

@article{li2019paradigm,
  title={{A New Paradigm for FPGA Placement Without Explicit Packing}}, 
  author={Li, Wuxi and Pan, David Z.},
  journal={IEEE Transactions on Computer-Aided Design of Integrated Circuits and Systems (TCAD)},
  volume={38},
  number={11},
  pages={2113-2126},
  year={2019},
  publisher={IEEE}
}

@inproceedings{lemieux2001using,
  title={{Using Sparse Crossbars within LUT}},
  author={Lemieux, Guy and Lewis, David},
  booktitle={ACM/SIGDA International Symposium on Field-Programmable Gate Arrays (FPGA)},
  pages={59--68},
  year={2001}
}

@inproceedings{shrivastava2025guaranteed,
  title={{Guaranteed Yet Hard to Find: Uncovering FPGA Routing Convergence Paradox}},
  author={Shrivastava, Shashwat and Nikoli{\'c}, Stefan and Tanaka, Sun and Ravishankar, Chirag and Gaitonde, Dinesh and Stojilovi{\'c}, Mirjana},
  booktitle={IEEE International Symposium on Field-Programmable Custom Computing Machines (FCCM)},
  pages={143--151},
  year={2025}
}

@misc{valgrind,
  title = {Valgrind},
  howpublished = {\url{https://valgrind.org}},
  note = {Accessed: 2026-01-17}
}

@inproceedings{luu2014towards,
  title={{Towards Interconnect-Adaptive Packing for FPGAs}},
  author={Luu, Jason and Rose, Jonathan and Anderson, Jason},
  booktitle={ACM/SIGDA International Symposium on Field-Programmable Gate Arrays (FPGA)},
  pages={21--30},
  year={2014}
}

@phdthesis{haroldsen2017academic,
    author={Haroldsen, Travis D.},
    title={{Academic Packing for Commercial FPGA Architectures}},
    school={Brigham Young University},
    year={2017}
}
